# Deep-glassy ice VI confirmed with a combination of neutron spectroscopy and diffraction

*Alexander Rosu-Finsen,[a] Alfred Amon,[a] Jeff Armstrong,[b] Felix Fernandez-Alonso[c,d,e] and Christoph G. Salzmann*[a]*

[a] Department of Chemistry, University College London, 20 Gordon Street, London WC1H 0AJ, United Kingdom.

[b] ISIS Pulsed Neutron and Muon Source, Rutherford Appleton Laboratory, Harwell Oxford, Didcot OX11 0QX, United Kingdom.

[c] Department of Physics and Astronomy, University College London, Gower Street, London, WC1E 6BT, United Kingdom.

[d] Materials Physics Center, CSIC-UPV/EHU, Paseo Manuel Lardizabal 5, 20018 Donostia - San Sebastian, Spain.

[e] IKERBASQUE, Basque Foundation for Science, Maria Diaz de Haro 3, 48013 Bilbao, Spain.

**Corresponding Author**

* c.salzmann@ucl.ac.uk

ABSTRACT

The recent discovery of a low-temperature endotherm upon heating hydrochloric-acid doped ice VI has sparked a vivid controversy. The two competing explanations aiming to explain its origin range from a new distinct crystalline phase of ice to deep-glassy states of the well-known ice VI. Problems with the slow kinetics of deuterated phases have been raised, which we circumvent here entirely by simultaneously measuring the inelastic neutron spectra and neutron diffraction data of $H_2O$ samples. These measurements clearly confirm the deep-glassy ice VI scenario and rule out alternative explanations. Additionally, we show that the crystallographic model of $D_2O$ ice XV, the ordered counterpart of ice VI, also applies to the corresponding $H_2O$ phase. The discovery of deep-glassy ice VI now provides a fascinating new example of ultra-stable glasses which are encountered across a wide range of other materials.

**TOC GRAPHICS**

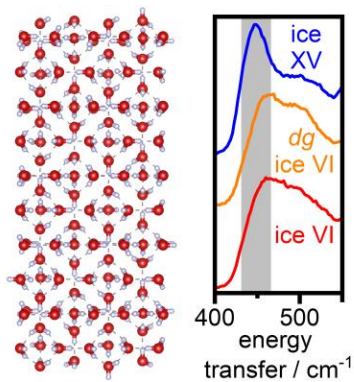

Ice is a highly complex material with currently 18 different polymorphs and at least two distinct amorphous forms identified.[1-2] The highly polar water molecule is capable of forming extended hydrogen-bonded networks with a wide range of different topologies. For a given network, the orientations of the hydrogen-bonded water molecules can adopt, in principle, all configurations from random orientational disorder to fully ordered structures with translation symmetry.[1] The exothermic phase transitions from so-called hydrogen-disordered phases to their hydrogen-ordered counterparts are expected to take place upon cooling. However, molecular reorientation processes in ice are defect-mediated and hence highly cooperative, which means that orientational glasses can be obtained upon cooling instead of hydrogen-ordered ices as the reorientation dynamics 'freeze-in'. This is illustrated, for example, by the differential scanning calorimetry (DSC) scan (1) of hydrogen-disordered ice VI in Figure 1(a) which shows a glass transition associated with the unfreezing of molecular reorientations upon heating at ~134 K.[3]

A major break-through in achieving hydrogen ordering of some of the high-pressure ice phases and, thus, preventing the formation of orientational glasses was the addition of hydrochloric acid (HCl) as a dopant, leading to the discovery of the hydrogen-ordered ices XIII, XIV and XV.[4-5] Achieving hydrogen ordering of the various phases of ice is important in order to clarify if their energetic ground states are ferro- or antiferroelectric,[6] make condensed phases of water available with strictly defined spectroscopic selection rules,[7] provide benchmark structures for testing the computer models of water,[6, 8-11] and understand geological processes in icy moons and planets.[12]

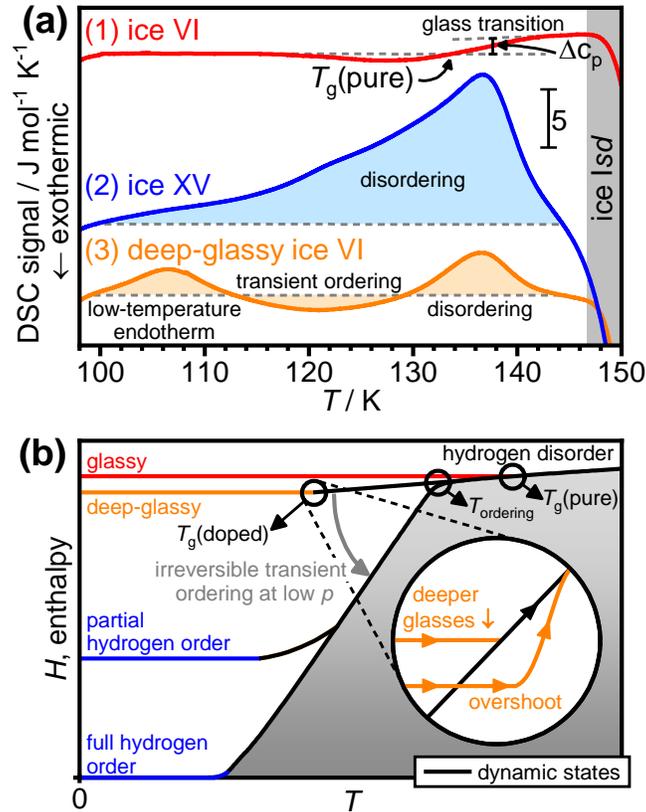

**Figure 1.** Calorimetry and thermodynamic pathways of pure and HCl-doped ice VI/XV samples. (a) Differential scanning calorimetry scans recorded at 10 K min$^{-1}$ at ambient pressure of (1) pure ice VI quenched at 1.0 GPa, (2) ice XV obtained after slow-cooling HCl-doped ice VI from 138 K at 0.2 K min$^{-1}$ at ambient pressure and (3) deep-glassy HCl-doped ice VI slow-cooled at 1.7 GPa at 0.25 K min$^{-1}$. (b) Schematic plot of enthalpy against temperature highlighting the various pathways pure and HCl-doped ice VI/XV can follow. The inset shows a magnification of the region around $T_g$(doped) where deep-glassy ice VI can take different pathways upon heating depending on the state of relaxation.

The formation of ice XV, the hydrogen-ordered counterpart of ice VI, has proven to be one of the most complex cases so far.[5-6, 9-11, 13-16] Much of the complexity arises from the fact that ice VI/XV consists of two interpenetrating hydrogen-bonded networks.[17-19] The formation of ice XV therefore goes along with the ordering of the individual networks as well as 'inter-network' ordering which ultimately determines its space group symmetry.[5, 19-20] Intriguingly, the formation of ice XV is accompanied by an increase in volume which is due to an expansion in the crystallographic $c$ direction.[19] This means that the formation of ice XV is fastest at low pressures.[5, 14, 20] A consequence of this is that pressure-quenched samples display an exothermic transient-ordering feature to ice XV upon heating at ambient pressure followed by the disordering phase transition to ice VI.[19-20] DSC scan (2) in Figure 1(a) shows the heating of a highly ordered ice XV sample at ambient pressure that has been obtained by slow-cooling at 0.2 K min$^{-1}$ from 138 K.[14, 20] Figure 1(b) shows a summary of the various possible thermodynamics pathways including the formation of 'deep-glassy' ice VI if hydrogen ordering can be suppressed, for example, by cooling under pressure.[21]

Cooling HCl-doped ice VI/XV at pressures greater than ~1.4 GPa leads to the appearance of a low-temperature endothermic feature before the onset of the transient ordering (*cf.* scan (3) in Figure 1(a)).[21-23] This endotherm has been assigned to the phase transition from a new hydrogen-ordered phase to ice XV.[22] According to Gasser *et al.*, their 'β-ice XV' is more hydrogen-ordered, differently hydrogen-ordered and more stable than ice XV.[22] Since the new endotherm is irreversible, we pointed out that it is impossible to calculate entropy changes from it and hence to derive direct structural information in terms of hydrogen order.[21] A major misunderstanding in ref. 22 is also that the ice XV to ice VI phase transition does not begin at 129 K but just above 100 K as shown in Figure 1(a) and elsewhere.[19-20] In contrast to the β-ice XV scenario, we argued that the low-temperature endotherm arises from the glass transition of deep-glassy ice VI which becomes increasingly more relaxed or 'deeper' as the cooling pressure is increased.[21] The cooling rate has only a minor effect. Deep-glassy states are well-known to produce endothermic

'overshoot' effects[24-26] upon heating as shown schematically in the inset of Figure 1(b). Accordingly, fast heating rates for a 1.0 GPa cooled sample and prolonged annealing below $T_g$(doped) at ambient pressure produced the new endotherm as well, and the deep-glassy ice VI scenario is also consistent with the X-ray diffraction and dielectric spectroscopy data in ref. 22.[21] Furthermore, we showed with neutron diffraction that a $D_2O$ sample that displayed a low-temperature endotherm has the structure of hydrogen-disordered ice VI.[21] However, it was subsequently argued that the formation of $D_2O$ β-ice XV may be hindered for kinetic reasons.[27]

Here we circumvent the potential complications associated with the $D_2O$ phases entirely, and carry out a combined inelastic neutron spectroscopy and neutron diffraction study of $H_2O$ samples. In this context, we discuss recently published Raman spectra presented in favor of the β-ice XV scenario.[27]

For this study, two different $H_2O$ ice VI/XV samples were prepared: Pure ice VI was obtained by pressure-quenching ice VI at 1.0 GPa and deep-glassy ice VI by slow-cooling HCl-doped ice VI at 2.5 K min$^{-1}$ at 1.7 GPa.[21-22] The inelastic neutron-scattering (INS) spectra of both samples were then measured at 15 K on the TOSCA instrument at ISIS[28] where the deep-glassy ice VI was also transformed to ice XV by heating to 138 K and subsequent cooling at 0.2 K min$^{-1}$. TOSCA is an indirect neutron spectrometer whose geometry has been specifically optimized for the study of hydrogen motions.[28] A recent upgrade of this instrument has provided up to a two-order of magnitude increase in count rates, thereby enabling parametric neutron-spectroscopic studies alongside simultaneous neutron diffraction.

Figure 2 shows the recorded INS spectra in the low energy-transfer range where modes associated with hindered translations and rotations (librations) of rigid water molecules are observed. It is well-known that this spectral range is particularly sensitive for highlighting the spectroscopic differences between hydrogen-disordered ices and their hydrogen-ordered counterparts.[7, 29-30] The recorded spectrum of ice VI is consistent with previous reports in the literature[31-32] and significantly different compared to the ice XV spectrum which is reported here for the first time. Compared to ice VI, ice XV displays additional sharp features at 168 and 448 cm$^{-1}$, and a broad feature at ~715 cm$^{-1}$. The gain in intensity at the librational edge for ice XV is consistent with an earlier Raman spectroscopic study[7] and has also been seen in the INS spectra of the ice VII/VIII disorder-order pair.[33]

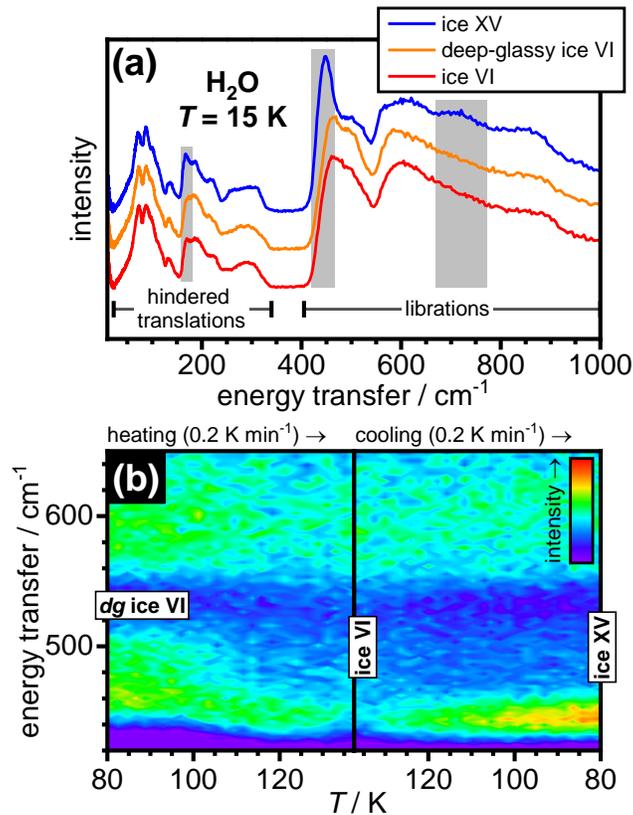

**Figure 2.** Inelastic neutron spectra of $H_2O$ ice VI/XV samples. (a) Spectra of ice VI, deep-glassy ice VI and ice XV collected at 15 K. The gray-shaded areas highlight the spectral range where major differences between the ice VI samples and ice XV were observed. The spectra are shifted

vertically for clarity. The entire available spectroscopic range is shown in Figure S1. (b) Contour plot of the librational region upon heating deep-glassy ice VI from 80 to 138 K, followed by cooling back to 80 K.

The characteristic ice XV spectroscopic features are missing entirely from the spectrum of deep-glassy ice VI, whose spectrum is very similar to that of ice VI. This confirms that the new low-temperature endotherm is indeed associated with the glass transition of deep-glassy ice VI, which is structurally very close to 'standard' ice VI. An INS spectrum of β-ice XV, which has been suggested to be more and differently ordered than ice XV,[22] would have to display a spectrum markedly different from both ice VI and ice XV.[7]

Figure 2(b) shows INS spectra collected upon heating deep-glassy ice VI from 80 to 138 K and cooling back at 0.2 K min$^{-1}$. Overall, the intensities of the spectroscopic features decrease upon heating due to the effect of temperature on the Debye-Waller factors. Nevertheless, the transient ordering can be seen from the appearance of the ~448 cm$^{-1}$ ice XV feature above 100 K and its subsequent disappearance at 138 K. Slow-cooling back to 80 K leads to significant increases in intensity of this feature illustrating, in line with earlier studies,[19-20] that slow-cooling produces a much more ordered ice XV compared to the states accessible during the transient ordering upon heating which is also consistent with the schematic shown in Figure 1(b).

In addition to the INS data, we also analyzed the diffraction data collected synchronously on the TOSCA instrument. Neutron diffraction (ND) of protiated samples generally suffers from intense background signals due to the strong incoherent scattering properties of $^1$H.[34] However, the coherent scattering cross section of $^1$H is sufficiently large compared to oxygen so that the diffraction pattern should contain information about the structure of the hydrogen sublattice.[34] The ND patterns of ice VI, deep-glassy ice VI and ice XV at 15 K are shown in Figure 3.

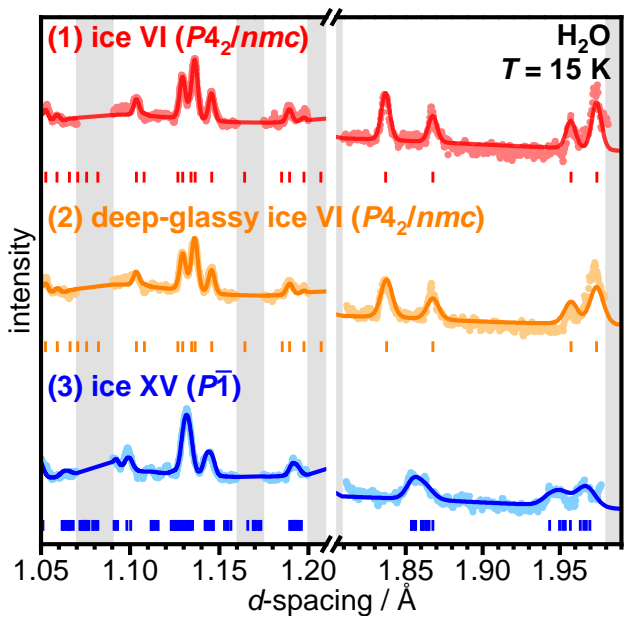

**Figure 3.** Neutron diffraction patterns of $H_2O$ (1) ice VI, (2) deep-glassy ice VI and (3) ice XV collected at 15 K. The experimental diffraction data are shown as light data points and the Rietveld fits as darker solid lines. Tick marks indicate the expected positions of Bragg peaks. Gray-shaded areas highlight excluded regions where strong diffraction features from the Al sample holder were observed. The Rietveld fits over the entire available $d$-spacing range are shown in Figure S2.

The ND data of ice VI and deep-glassy ice VI are remarkably similar and both datasets could be fitted well using the $P4_2/nmc$ structural model of ice VI. The ND pattern of ice XV on the other hand is very different and lacks, for example, a Bragg peak at ~1.84 Å characteristic of the two disordered phases. The Rietveld analysis of the ice XV pattern was performed using the $P$-1 crystallographic model derived earlier for the corresponding $D_2O$ phase.[5, 19] The ND data at low

*d*-spacing can be regarded as a 'fingerprint region' due to the many overlapping Bragg peaks. The good agreements in this part of the ND patterns therefore underpin the validities of the used crystallographic models. More detailed information on the Rietveld analyses is given in the Supporting Information. So far, the crystal structures of the hydrogen-ordered ices were derived from $D_2O$ phases, and it has been generally assumed that the corresponding $H_2O$ phases hydrogen-order in similar fashions. The ND data of $H_2O$ ice XV presented here provides the first evidence for this assumption. Furthermore, we note that the proposed β-ice XV, which has been suggested to have a different space group symmetry compared to ice XV,[22] would have to display a significantly different ND pattern compared to the one of deep-glassy ice VI and that of ice XV.

In summary, it is clear that the 'standard' and deep-glassy ice VI are very similar from both the spectroscopic as well as the diffraction point of view, and both are very different compared to the hydrogen-ordered ice XV. These observations rule out the β-ice XV scenario[22] and firmly link the low-temperature endotherm with deep-glassy states of ice VI.

Despite the strong similarities between 'standard' and deep-glassy ice VI, our scenario requires the two states to be slightly different in enthalpy (*cf*. Figure 1(b)). In ref. 21 we argued that this enthalpy difference could be realized with very local hydrogen-ordered domains that do not display any orientational correlations between them, so that the average structure is fully hydrogen disordered. Such small domains would then cause strain on the surrounding disordered matrix. Indeed, the Bragg peaks of deep-glassy ice VI are broader compared to 'standard' ice VI[21] and this effect is also seen for the Bragg peaks at higher *d*-spacing in Figure 3. Taken to the extreme, it could be argued that such local distortions could even degrade the space group symmetry of ice VI. At this point it is important to recall that all hydrogen-disordered phases of ice display strictly speaking only *P*1 space group symmetry due to the lack of translational symmetry.[19]

Compared to ice VI, the Raman spectra of what we have assigned to deep-glassy ice VI display small shifts in the O-H stretching modes and the high-wavenumber component in the decoupled O-D stretching mode region seems to become more Raman active.[22, 27] Changes in the relative intensities of modes in micro-Raman spectroscopy can arise from preferred orientation effects[35] which clearly play a role as can be seen from the comparison of nominally identical ice VI/XV materials in the literature.[7, 22, 27, 29, 36-38] However, the shift in the coupled O-H stretching mode region may be a sensitive indicator for the strain levels in deep-glassy ice VI. In fact, Raman spectroscopy has been demonstrated to be a very powerful technique for detecting strain in a wide range of different materials.[39-41] Since Raman intensities arise from changes in the polarizabilities, it can in principle be more sensitive to strain compared to INS.

The identification of deep-glassy states of ice now adds a fascinating new facet to ice research. In fact, it seems as if deep-glassy states and the associated phenomenon of transient ordering have just recently been identified for the 'ordinary' ice I*h* as well.[42] Beyond ice, deep-glassy states of natural amber have recently been identified[26] and the vapor-deposition technique has been shown to be particularly suited for preparing ultra-stable glasses with a wide range of scientific challenges and future applications anticipated.[43] Despite the similar underlying thermodynamics, linking the 'deepness' of the glassy states with the details of the local structure will be a highly individual challenge for each material in question.

**Experimental Methods**

Pure $H_2O$ ice VI samples were obtained by quenching ice VI at 1.0 GPa to 77 K with liquid nitrogen in a piston-cylinder setup, whereas deep-glassy ice VI was prepared by slow-cooling 0.01 HCl-doped $H_2O$ ice VI at 1.7 GPa at 2.5 K min$^{-1}$ to 77 K. Detailed descriptions of the sample preparations can be found in refs 5, 7, 14, 19-21. DSC scans of the samples were recorded using a Perkin Elmer DSC 8000 as described previously.[14, 20-21] About 3 g of both samples were ground using a porcelain pestle and mortar under liquid nitrogen, and transferred into precooled flat Al cans with a sample-compartment thicknesses of 2 mm. The sample cans were then mounted onto the cryostat sticks of the helium cryostat of the TOSCA indirect neutron spectrometer at ISIS.[28] After quick transfers into the cryostat, both samples were measured at 15 K for at least 2 hours. In case of the deep-glassy ice VI sample, this was followed by heating from 80 K to 138 K at 0.2 K min$^{-1}$ and cooling back to 80 K at 0.2 K min$^{-1}$ to give ice XV while continuously collecting spectra in 6-minute runs. The scattering from the Al can in the ±45° spectroscopy banks is negligible in comparison to ice. TOSCA also contains diffraction detectors in back-scattering position at 179°. The ND patterns were analyzed using the GSAS program[44] as described in detail in the Supporting Information.

ASSOCIATED CONTENT

Supporting Information containing additional INS and ND data as well as details on the Rietveld refinements.

AUTHOR INFORMATION

**Notes**

The authors declare no competing financial interests.

ACKNOWLEDGMENT

We thank the Royal Society for a University Research Fellowship (CGS, UF150665), the Austrian Science Fund for a Schrödinger fellowship (AA) and ISIS for granting access to the TOSCA instrument. Furthermore, this project has received funding from the European Research Council (ERC) under the European Union's Horizon 2020 research and innovation program (grant agreement no. 725271).

# Supporting Information

# Deep-glassy ice VI confirmed with a combination of neutron spectroscopy and diffraction

*Alexander Rosu-Finsen,[a] Alfred Amon,[a] Jeff Armstrong,[b] Felix Fernandez-Alonso[c,d,e] and Christoph G. Salzmann*[a]*

[a] Department of Chemistry, University College London, 20 Gordon Street, London WC1H 0AJ, United Kingdom.

[b] ISIS Pulsed Neutron and Muon Source, Rutherford Appleton Laboratory, Harwell Oxford, Didcot OX11 0QX, United Kingdom.

[c] Department of Physics and Astronomy, University College London, Gower Street, London, WC1E 6BT, United Kingdom.

[d] Materials Physics Center, CSIC-UPV/EHU, Paseo Manuel Lardizabal 5, 20018 Donostia - San Sebastian, Spain.

[e] IKERBASQUE, Basque Foundation for Science, Maria Diaz de Haro 3, 48013 Bilbao, Spain.

## Contents

## 1. Full-range INS data

Figure S1 shows the full spectral range of the inelastic neutron spectra shown in Figure 2 in the main article. The various modes have been assigned following the computational normal-mode analysis presented in ref. 1.

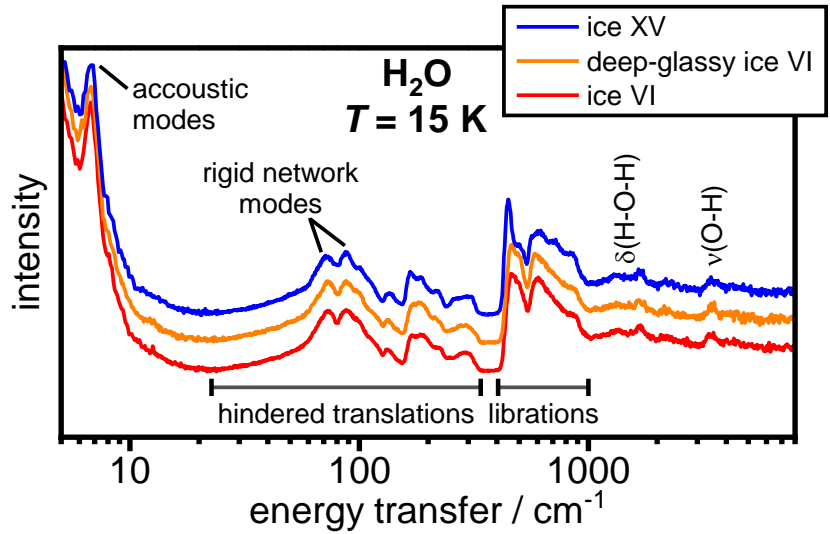

***Figure S1.*** *Inelastic neutron spectra of $H_2O$ ice VI, deep-glassy ice VI and ice XV collected at 15 K on the TOSCA instrument at ISIS. The spectra are shifted vertically for clarity. Note the logarithmic scale of the abscissa.*

## 2. Full-range neutron diffraction data and Rietveld refinements

In addition to the diffraction data shown in Figure 3 in the main article, Figure S2 shows the Rietveld fits across the *d*-spacing range presently available on the instrument. The crystallographic models of ice VI/XV were taken from the corresponding $D_2O$ samples in refs 2 and 3 with *P4$_2$/nmc* and *P*-1 space group symmetries, respectively. The Rietveld refinements included the optimisation of the lattice constants, scale factor, peak profile parameters, background function (twelve-term shifted Chebyschev), and thermal displacement parameters of the protium and oxygen atoms which were constrained to take the same values for a given type of atom. The fractional occupancies of the hydrogen positions were set to 0.5 for the $H_2O$ ice VI samples whereas the fractional occupancies from the corresponding $D_2O$ ice XV sample (cooled at 0.07 K min$^{-1}$) were used for the $H_2O$ ice XV.[3] Judging from the calorimetric data shown in Figure 4 of ref. 4, the difference in the fraction of Pauling entropy between $H_2O$ ice XV (cooled at 0.2 K min$^{-1}$) and $D_2O$ ice XV (cooled at 0.07 K min$^{-1}$) is <2% which means that the fractional occupancies of the $D_2O$ sample from ref. 3 should be very close to the ones of the corresponding $H_2O$ sample.

The *d*-spacing values of the collected diffraction data were calibrated using the strong Bragg peaks from the Al sample can and the lattice constant of Al at 20 K from the literature (4.0322 Å).[5-7]

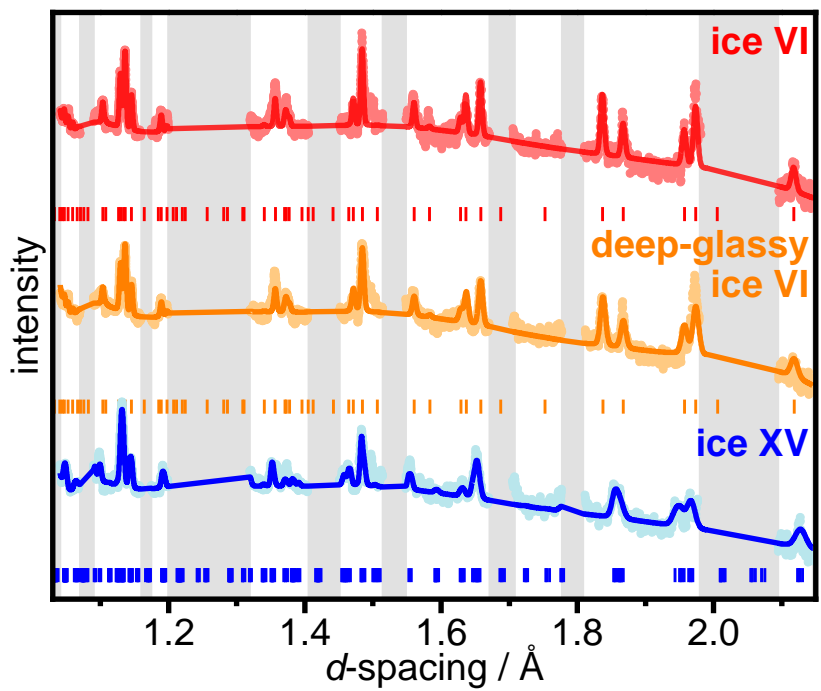

*Figure S2.* Neutron diffraction patterns of $H_2O$ ice VI, deep-glassy ice VI and ice XV collected at 15 K on the TOSCA instrument. The experimental diffraction data are shown as light data points and the Rietveld fits as darker solid lines. Tick marks indicate the expected positions of Bragg peaks. Gray-shaded areas indicate excluded regions where strong diffraction features from the sample holder were observed.

The refined lattice constants of the three $H_2O$ ice VI/XV samples at 15 K are given in Table S1. Within the margins of error, the lattice constants of the pure and the deep-glassy ice VI samples are indistinguishable. However, it is important to note that the changes in the lattice constants upon going from the $H_2O$ ice VI samples to $H_2O$ ice XV follow the same trends as previously observed for the corresponding $D_2O$ samples:[3] The *a* and *b* lattice constants contract upon hydrogen ordering whereas a significant expansion in the *c* parameter is observed. Judging from the volumes of the unit cells,[3] hydrogen-ordered H$_2$O ice XV is less dense than the hydrogen-disordered H$_2$O ice VI samples at 15 K and ambient pressure. The same relative density difference has also been observed for the corresponding D$_2$O samples.[3]

***Table S1.*** *Refined lattice constants of H$_2$O ice VI, deep-glassy ice VI and ice XV using diffraction data collected at 15 K on the TOSCA instrument. The structural models were based on the P4$_2$/nmc model of ice VI[2, 8] and the P-1 model of ice XV.[3, 9] The numbers in parentheses indicate the errors of the last significant figure.*

| sample | $a$ / Å | $b$ / Å | $c$ / Å | $\alpha$ / ° | $\beta$ / ° | $\gamma$ / ° | $V$ / Å$^3$ |
|---|---|---|---|---|---|---|---|
| ice VI | 6.2425(4) | 6.2425(4) | 5.7667(5) | 90 | 90 | 90 | 224.72(3) |
| deep-glassy ice VI | 6.2421(5) | 6.2421(5) | 5.7686(8) | 90 | 90 | 90 | 224.77(4) |
| ice XV | 6.2262(7) | 6.2105(6) | 5.8303(6) | 89.90(2) | 89.85(2) | 89.89(1) | 225.44(4) |